\documentclass[aps,twocolumn]{revtex4}
\usepackage[utf8]{inputenc}
\usepackage{graphicx}
\begin{document}
\title{Ensemble velocity of non-processive molecular motors with multiple
  chemical states} \author{Andrej Vilfan} 
\email{andrej.vilfan@ijs.si}
\affiliation{J. Stefan Institute, Jamova 39, 1000 Ljubljana, Slovenia}
\affiliation{Faculty of Mathematics and Physics, University of Ljubljana,
  Jadranska 19, 1000 Ljubljana, Slovenia}
\begin{abstract}We study the ensemble velocity of non-processive motor
  proteins, described with multiple chemical states. In particular, we discuss
  the velocity as a function of ATP concentration.  Even a simple model which
  neglects the strain-dependence of transition rates, reverse transition rates
  and nonlinearities in the elasticity can show interesting functional
  dependencies, which deviate significantly from the frequently assumed
  Michaelis-Menten form.  We discuss how the oder of events in the duty cycle
  can be inferred from the measured dependence. The model also predicts the
  possibility of velocity reversal at a certain ATP concentration if the duty
  cycle contains several conformational changes of opposite directionalities.
  \end{abstract}
\maketitle  

\section{Introduction}

Motor proteins are molecular machines that convert chemical energy, usually
obtained from the hydrolysis of ATP, to mechanical work by walking along their
tracks \cite{Howard_book,Kolomeisky.Fisher2007}. They can be classified as
processive and non-processive \cite{howard97}. Processive motors have the
ability to make many steps before detaching from the track and a single motor
molecule is sufficient to transport a cargo over a significant distance. In
most cases processive motors are dimeric and alternately move their heads in a
hand-over-hand fashion. Notable examples include most kinesins, myosin V,
myosin VI and cytoplasmic dyneins. Non-processive motors dissociate from the
track after each step, but can still move loads over long distances when
cooperating in large numbers. Non-processivity is usually connected with a low
duty ratio - the motor spends a large fraction of the cycle detached from the
track. Best known non-processive motors are muscle myosins and axonemal
dyneins. Many processive motors become non-processive in the monomeric form
\cite{hancock98}.

A number of studies has been devoted to the velocity of processive motors as a
function of load and ATP concentration. For a processive motor one expects and
finds that both the ATP hydrolysis rate and the velocity of the motor follow
the Michaelis-Menten dependence on the ATP concentration
\cite{howard89,Steinberg.Schliwa1996,schnitzer97,visscher99}
  \begin{equation}
\label{eq:michaelis}
v=\frac{v_{\rm max}\rm [ATP]}{K_m+\rm [ATP]}\;.
\end{equation}
The load dependence is more complex.  Duke and Leibler proposed that some
properties of kinesin's force velocity relation could be explained even
without any coordination of the two chemical cycles \cite{duke96}.
Alternatively, some models assume tight coordination and use the load
dependence to construct diagrams with several states and mechanical substeps
\cite{Fisher.Kolomeisky2001,Tsygankov.Fisher2007}.

In non-processive motors the situation is fundamentally different. Because
each motor spends a significant part of its cycle in the detached state, the
distance travelled per ATP hydrolyzed is not simply related to the step
size. In fact, different estimates
\cite{Toyoshima.Spudich1990,Uyeda.Spudich1990,Harada.Yanagida1988} of the
distance per step led to a long lasting controversy about the myosin
mechanism. Nevertheless, many models for muscle myosin were developed, aimed
at relating the velocity or transient response of muscle sarcomeres to the
properties of a single myosin molecule
\cite{huxley57,hill74,eisenberg80}. They were able to predict the principal
features of the myosin cycle long before any structural evidence was
available. The aforementioned models studied the limit of a large number of
motors and assumed that the filaments are sliding at a constant velocity. In
other words, they excluded the possibility that the cycles of motors become
correlated and produce non-uniform motion. This possibility was explored by
Duke \cite{duke99,duke2000} who showed that a filament under high load can
indeed show a synchronization of chemical cycles between myosin heads and
step-wise motion. A further complication arises from the structure of a muscle
fibril with many sarcomeres in series. This can cause spontaneous symmetry
breaking and individual motors can be subject to a different stretch than the
macroscopic sarcomere \cite{vilfan2003b}. The application of an abrupt force
step can transiently synchronize the motors and lead to observable
oscillations \cite{Edman.Curtin2001}, as predicted by Duke's model.

\begin{figure*}
  \begin{center}
    \includegraphics[width=0.65\textwidth]{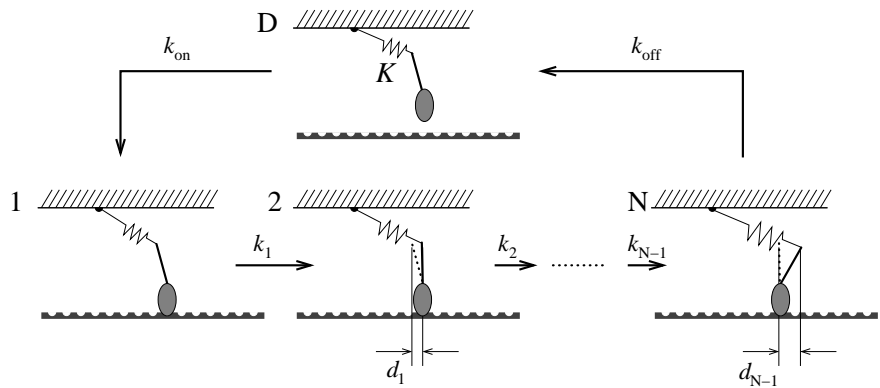}
  \end{center}
  \caption{Duty cycle of a motor, consisting of $N$ bound states with distinct
    lever orientations and irreversible transitions between them. A detached
    motor (state D) binds to the track with rate $k_{\rm on}$, which leads to
    state 1 and the elastic element is initially unstrained. The next
    transition ($1\to 2$) takes place with the rate $k_1$ and includes a lever
    arm movement of distance $d_1$. Eventually, the motor detaches from state
    $N$ with the rate $k_{\rm off}$.}
  \label{figure1}
\end{figure*}

Less attention has been paid to the velocity of non-processive motors as a
function of the ATP concentration. It is frequently assumed to follow a
Michaelis-Menten like dependence, as would be the case with processive motors,
even though there is no reason why it should have that form. Some experimental
studies on myosins either show no deviation from the Michaelis-Menten shape
\cite{kron-spudich:86,Warshaw.Trybus1991,Debold.Warshaw2008,Debold.Walcott2011,Canepari.Bottinelli2012,Persson.Mansson2013},
whereas others show minor, but significant deviation \cite{Baker.Warshaw2002}.
Axonemal dyneins also largely follow the Michaelis-Menten dependence
\cite{Kagami.Kamiya1992,Lorch.Hunt2008}.  An expression for a specific model
has been discussed in the seminal paper by Leibler and Huse
\cite{leibler93}. They show that the dependence is described by a more complex
function, which they call generalized Michaelis-like law. However, the
dependence still has a similar functional form -- as we will see this is
related to the assumption that the power stroke takes place as the next step
after binding.

The aim of this paper is to discuss the dependence $v(\rm [ATP])$ in a more
general context. The study is motivated by other non-processive motors that do
not necessarily have a power-stroke right after binding. We can mention
non-processive kinesins \cite{Cross.McAinsh2014} (e.g., kinesin-14 or ncd
\cite{Cross2010}) as examples of such motors. Because the ATP dependence of
the velocity can be measured in a relatively simple motility assay
\cite{duke95}, our model should provide a way to extract some properties of
the duty cycle that are difficult to measure in a single-molecule experiment.

\begin{figure*}
  \begin{center}
    \includegraphics[width=0.8\textwidth]{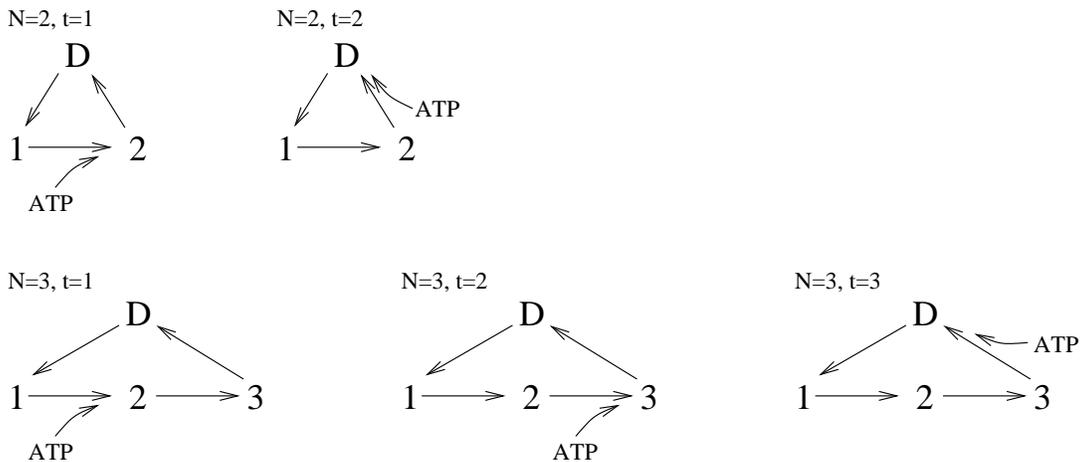}
  \end{center}
  \caption{Cycles with $N=2$ bound states (upper row) and $N=3$ bound states
    (lower row).  The cycles are classified depending on whether ATP binding
    occurs during the transition $1\to 2$ ($t=1$), $2\to \rm D (3)$ ($t=2$) or
    $\rm 3 \to D$ ($t=3$). }
  \label{figure2}
\end{figure*}

\section{Model}

In the following we will discuss a generalization of the ``rower'' model
\cite{leibler93} by introducing several substeps. On the other hand, we still
make a number of simplifying assumptions, notably that all elastic elements
are linear, that the chemical transitions are irreversible and their rates
independent of strain.  We further assume that the motors always run through
the same cycle, which involves hydrolysis of 1 ATP molecule. All these
simplifications are limited in their validity, but should work well in a
motility assay where forces are relatively small.  As a rule of thumb, the
strain dependence of transition rates can be neglected if the elastic energy
change during a step is smaller than the thermal energy. This will be the case
if the motors are attached with their flexible tails to the glass surface. For
example, different single-molecule studies report myosin's elastic constant as
$0.7\,\rm pN/nm$ for full length HMM \cite{veigl98}, but $1.8\,\rm pN/nm$ for
the S1 fragment lacking the tail \cite{Lewalle.Sleep2008}. In a muscle
sarcomere the tails are tightly packed into thick filaments, which also
increases their effective stiffness \cite{Kaya.Higuchi2013}. The assumption of
linear elastic elements will also hold for relatively small forces that are
exerted by motors in a motility assay. Finally, neglecting reverse transitions
in the cycle is valid if there is no ADP and phosphate in solution (which
would lead to reversal of product release steps) and, again, if the forces are
too low for significant mechanically induced reverse transitions.  In a recent
paper Persson \textit{et al.} \cite{Persson.Mansson2013} discuss the role of
the aforementioned effects in a motility assay for rather stiffly anchored
heads ($2.8\,\rm pN/nm$) and show that off-path transitions and nonlinear
elasticity can be important, although the resulting dependence is close to the
Michaelis-Menten form. We further assume that the filaments (or tracks) are
straight and stiff, and that all motors interacting with them all act in the
same way and move the filament along its axis.

The motors are modelled as shown in Fig.~\ref{figure1}. We assume that each
motor is connected through an elastic element (spring constant $K$) to the
backbone (or surface in a motility assay). The motor is initially in the
dissociated state (state D). It binds to the track with the rate $k_{\rm
  on}$. After that it undergoes its first conformational change with a rate
$k_1$, which moves the lever by a distance $d_1$. The next step, taking place
with a rate $k_2$ moves it by $d_2$, ... until it reaches the last bound state
$N$ from where it detaches with a rate $k_{\rm off}$.  We denote the average
dwell times in those states with $\tau_1=1/k_1$, ..., $\tau_N=1/k_{\rm off}$
and $\tau_{\rm det}=1/k_{\rm on}$.  One of the steps requires binding of ATP
and is therefore a second-order transition. If this is the constant $k_t$, the
corresponding dwell time is $\tau_t=1/(k_t \rm[ATP])$.

In the above scheme 1 ATP molecule is always hydrolyzed per cycle and the
ATPase rate per motor can be expressed as
\begin{equation}
  \label{eq:atpase}
  v_{\rm ATPase}=\frac 1 {\tau_{\rm det}+\sum_{i=1}^N \tau_i} = \frac 1
  {\tau_{\rm det}+\sum_{i\ne t} \tau_i + 1/(k_t {\rm [ATP]}) }\;,
\end{equation}
which always follows the Michaelis-Menten dependence.

In the following we will derive the velocity of a large ensemble of motors
acting between the same backbone and track. In the limit of a large motor
number, $n\to \infty$, the motion becomes uniform and the velocity
fluctuations due to random transitions in individual motors negligible
(simulation results for finite sized systems are shown in Appendix
\ref{appendix}). We can examine the steady state of the system by assuming a
constant velocity $v$, then calculating the average force produced by each
motor and finally determining $v$ from force equilibrium.

With $\xi$ we denote the position of the lever arm relative to state 1, i.e.,
it has the values $\xi=0$ in state 1, $\xi=d_1$ in state 2, $\xi=d_1+d_2$ in
state 3, and so on. The ensemble average of $\xi$ among all bound motors is
then
\begin{equation}
  \label{eq:nlpos}
  \langle \xi \rangle = \frac{\sum_{i=2}^N \sum_{j=1}^{i-1} d_j \tau_i}{\sum_{i=1}^N\tau_i}\;.
\end{equation}
As the track is moving with velocity $v$, the average strain on a
spring is reduced by the expectation value of $\langle vt\rangle$ among bound
motors, where $t$ is the time passed since initial binding. It can be
calculated using
\begin{equation}
  \label{eq:texp}
  \langle t \rangle = \frac{\int_0^\infty P(t) t \, dt}{\int_0^\infty P(t) dt}
 = \frac{\sum_{i=1}^N  \int_0^\infty P_i(t) t \, dt }{\sum_{i=1}^N  \int_0^\infty P_i(t) dt }
\end{equation}
where $P(t)$ is the probability that the motor is still attached at time $t$
after its initial attachment and $P_i(t)$ is the probability that it is in
$i$-th bound state. The integrals are $\int_0^\infty P_i(t) dt=\tau_i$ and
$\int_0^\infty P_i(t) t\, dt=(\tau_1+\tau_2+\ldots+\tau_i)\tau_i$. Together,
these expressions give us the value
\begin{equation}
  \label{eq:strain}
  \langle x \rangle =   \langle vt \rangle =   v \frac{\sum_{i=1}^N \sum_{j=1}^i \tau_i
    \tau_j}{\sum_{i=1}^N \tau_i}\;.
\end{equation}

The total force produced by all motors can be expressed as
\begin{equation}
  \label{eq:force}
  F=\langle n \rangle K (\langle \xi \rangle - \langle x \rangle)
\end{equation}
with $n$ denoting the number of attached motors,
\begin{equation}
  n=n_{\rm tot} \left(1-\frac{\tau_{\rm det}}{\sum_{i=1}^N \tau_i +\tau_{\rm det}}\right)\;.
\end{equation}

In a gliding assay the friction is generally negligible and the force
equilibrium states $F=0$. From Eq.~(\ref{eq:force}) it follows that $\langle
\xi \rangle = \langle x \rangle$ and we obtain an expression for the
velocity
\begin{equation}
  \label{eq:velocity}
  v=\frac{\sum_{i=2}^N \sum_{j=1}^{i-1} d_j \tau_i}{\sum_{i=1}^N \sum_{j=1}^i \tau_i
    \tau_j}\;.
\end{equation}
In the following we will discuss the properties of this equation. 

\begin{figure*}
  \begin{center}
    \includegraphics[width=0.8\textwidth]{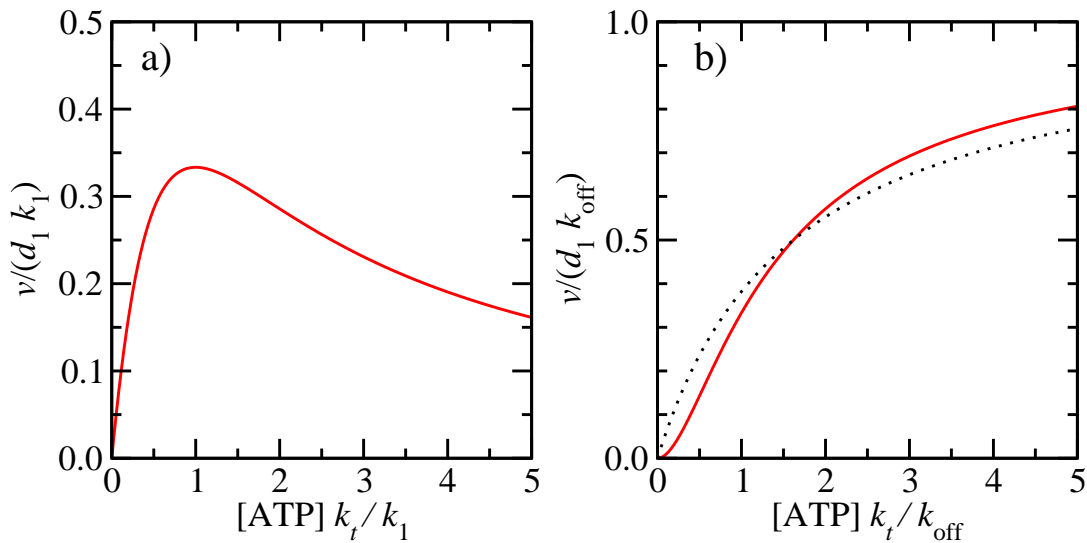}
  \end{center}
  \caption{Ensemble velocity as a function of the dimensionless ATP
    concentration for the model with two bound states. a) Model with ATP
    binding following the power stroke. b) ATP binding preceding power
    stroke. The dotted line shows the Michaelis-Menten dependence.}
  \label{figure3}
\end{figure*}

\subsection{Model with two bound states}

The simplest non-trivial case involves $N=2$ bound states (Fig.~\ref{figure2},
upper row): the first directly after binding and the second after a power
stroke of distance $d_1$.  The expression for velocity simplifies to
\begin{equation}
  \label{eq:v2}
  v=\frac{d_1 \tau_2}{\tau_1^2+\tau_1\tau_2+\tau_2^2}
\end{equation}
If ATP binds on the second transition ($t= 2$) we get the following
concentration dependence (Fig.~\ref{figure3}a):
\begin{equation}
  \label{eq:v2-1}
  v=\frac{d_1 k_t \rm [ATP] }{1+(k_t/k_1) {\rm [ATP]} + ( (k_t/k_1) {\rm [ATP]})^2}
\end{equation}
As expected, this dependence is linear at low ATP concentrations, but the
velocity somewhat surprisingly drops at high ATP concentrations, too. The
reason is that the post-powerstroke state becomes short lived and motors
produce forward force only for a small fraction of time. The maximum velocity
$v_{\rm max}=d_1 k_1 /3$ is achieved at ${\rm [ATP]}=k_1/k_t$.  The
non-monotonic dependence shows that the model is not adequate to describe the
properties of muscle myosin, which needs at least two post-powerstroke states
(ADP and rigor).

The second possibility is that ATP binding is the first step ($t=1$). Then the
expression for velocity is (Fig.~\ref{figure3}b):
\begin{equation}
  \label{eq:v2-2}
  v=\frac{d_1 k_{\rm off}  ((k_t/k_{\rm off})  \rm [ATP])^2}{1+(k_t/k_{\rm
      off}) {\rm [ATP]} + ( (k_t/k_{\rm off}) {\rm [ATP]})^2}
\end{equation}
The dependence is quadratic at low ATP concentrations. In
this regime the ensemble of bound motors is dominated by motors in the
pre-powerstroke state waiting for ATP binding. An increasing ATP concentration
not only accelerates the cycling rate, but also increases the fraction of
post-powerstroke motors in the ensemble, hence quadratic dependence. The
maximum velocity at saturating [ATP] is $v_{\rm max}=d_1 k_{\rm off}$ and half
the maximum is reached at ${\rm [ATP]}=\varphi \frac {k_{\rm off}}{k_t}$,
where $\varphi=(1+\sqrt{5})/2$ represents the golden ratio.

This simple model with 2 bound states demonstrates that the velocity shows a
dependence that is profoundly different from the frequently used
Michaelis-Menten like dependence. It also allows us to determine whether the
power stroke takes place before or after ATP binding.

\subsection{Model with three bound states}

In the following, we allow $N=3$ consecutive bound states (Fig.~\ref{figure2},
lower row). The lever movements are $d_1$ (between 1 and 2) and $d_2$ (between
2 and 3). Again, we have to distinguish between schemes where ATP binds to the
first, second or third state.

If ATP binding is the first transition ($t=1$), the expression for the
velocity reads
\begin{widetext}
\begin{equation}
\label{eq:3state1}
  v=\frac{d_1/k_2+(d_1+d_2)/k_{\rm off} }{1/(k_t{\rm [ATP]})^2+(1/k_2+1/k_{\rm
      off})/(k_t{\rm [ATP]}) +1/k_2^2+1/k_{\rm off}^2 + 1/(k_2 k_{\rm off}) }\;.
\end{equation}
The functional dependence of this equation is similar to
Eq.~(\ref{eq:v2-2}). The deviation is maximal when $k_2=k_{\rm off}$, but even
in this case the difference never exceeds $0.008\,v_{\rm max}$. 
(Fig.~\ref{figure4}a).

\begin{figure*}
  \begin{center}
    \includegraphics[width=0.8\textwidth]{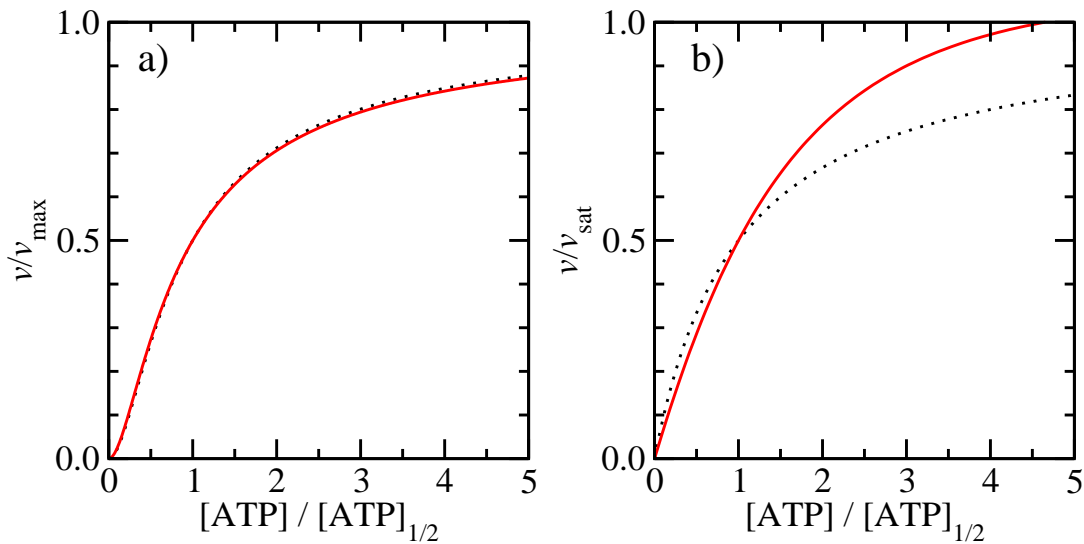}
  \end{center}
  \caption{Velocity in the model with 3 bound states. a)  ATP binding is the
    first transition, (\ref{eq:3state1}). The continuous line shows the case
    $k_2=k_{\rm off}$. The dotted line shows the result of the model with 2
    bound states (\ref{eq:v2-2}). b) ATP binding is the second step
    (\ref{eq:3state2}). The parameters were $k_1=k_{\rm off}$ and $d_2=0$. The
    dotted line shows the Michaelis-Menten dependence (\ref{eq:michaelis}).}
  \label{figure4}
\end{figure*}

The situation becomes different if ATP binding is the second step, $t=2$. Then
we get
\begin{equation}
\label{eq:3state2}
  v=\frac{d_1/(k_t{\rm [ATP]}) +(d_1+d_2)/k_{\rm off} }{1/(k_t{\rm [ATP]})^2+(1/k_1+1/k_{\rm
      off})/(k_t{\rm [ATP]}) +1/k_1^2+1/k_{\rm off}^2 + 1/(k_1\ k_{\rm off}) }
\end{equation}

In the special case $d_1=0$ the ATP waiting state becomes pre-powerstroke and
the functional form is the same as (\ref{eq:3state1}). We expect that this
scenario describes the cycle of single-headed kinesin, which binds to the
microtubule in ADP state, releases ADP, binds ATP and docks the neck linker
(which effectively represents a power stroke of distance $d_2$), hydrolyzes
ATP and detaches \cite{hoenger00,Guydosh.Block2009}.

Alternatively, for $d_2=0$ Eq.~(\ref{eq:3state2}) represents the
``Michaelis-like law'', as identified by Leibler and Huse \cite{leibler93}
(Fig.~\ref{figure4}b). At high ATP concentrations the velocity
approaches the saturation value faster and, depending on parameter values, may
also overshoot for the same reason as the two-state model
(Eq. \ref{eq:v2-2}).

Equation (\ref{eq:3state2}) becomes particularly interesting if $d_1$ and
$d_1+d_2$ have opposite signs (we assign $d_1<0$). This means that the motor
first binds, makes a conformational change in negative direction, waits for
ATP to bind and then commits the actual power stroke in the positive
direction.  The velocity then changes direction at a certain ATP
concentration. At low concentrations, the motors move in the direction of
$d_1$ and at high concentrations in the direction of $d_1+d_2$. Examples of
such curves are shown in Fig.~\ref{figure5}.

Finally, if ATP binding is the third step and triggers detachment ($t=3$), the
velocity is given by
\begin{equation}
\label{eq:3state3}
v=\frac{d_1/k_2 + (d_1+d_2)/(k_t{\rm [ATP]}) }{1/(k_t{\rm
    [ATP]})^2+(1/k_1+1/k_2)/(k_t{\rm [ATP]}) +1/k_1^2+1/k_2^2
  + 1/(k_1\ k_2) }\;.
\end{equation}
\end{widetext}

\begin{figure}
  \begin{center}
    \includegraphics[width=0.4\textwidth]{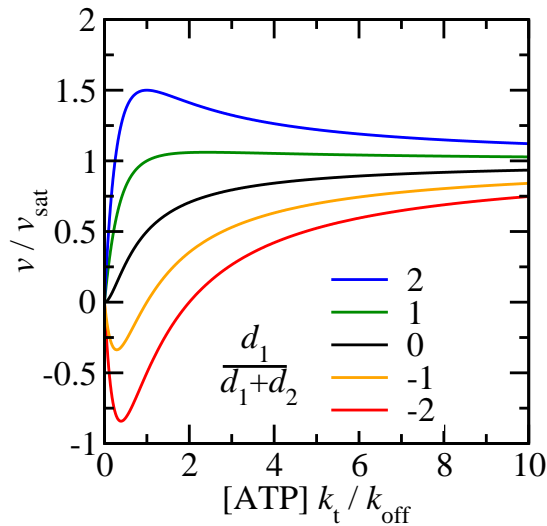}
  \end{center}
  \caption{Velocity as a function of the ATP concentration in the model with
    three bound states and ATP binding as the second transition. Depending on
    the ratio $d_1/(d_1+d_2)$, the dependence can be non-monotonic and even
    show a velocity reversal. All graphs are for $k_1=k_{\rm off}$.}
  \label{figure5}
\end{figure}

The properties of this expression are similar to the case when ATP binding is
the second step. We expect that this expression should describe the dependence
for muscle myosin (myosin II), whose main bound states are A.M.ADP.Pi, A.M.ADP
and A.M \cite{lymn71,Pate.Cooke1989a}. The main power stroke ($d_1$) takes
place along with phosphate release, but there is a second, smaller
conformational change upon ADP release ($d_2$). Single molecule experiments
give values of $d_1\approx 5\,\rm nm$ (which is possibly an underestimate
\cite{Sleep.Smith2006}) and $d_2\approx 1-2\,\rm nm$
\cite{Veigel.KendirckJones2003,Capitanio.Bottinelli2006}.

  \begin{figure}
    \begin{center}
      \includegraphics[width=0.35\textwidth]{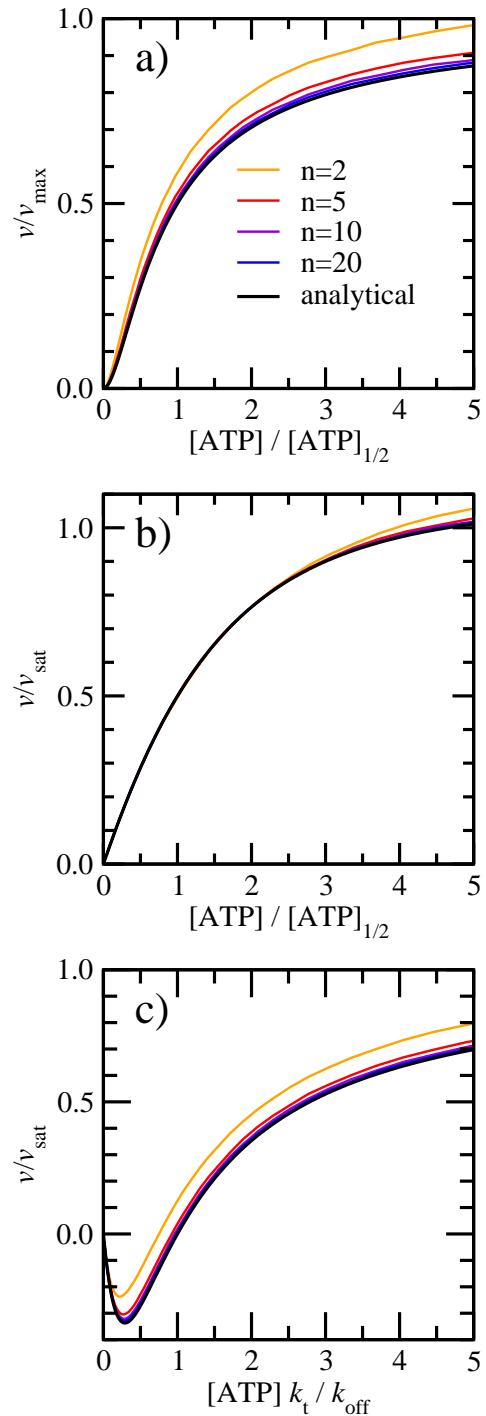}
    \end{center}
    \caption{Simulation results on systems of $n=2,\ldots,20$ motors, compared
      to the analytical result for an infinite ensemble.  a) System from
      Fig.~\ref{figure4}a with $N=3$ bound states, ATP binding as first
      transition ($t=1$) and $k_2=k_{\rm off}$. b) System from
      Fig.~\ref{figure4}b with $N=3$, $t=2$, $d_2=0$ and $k_1=k_{\rm off}$. c)
      System from Fig.~\ref{figure5} with $N=3$, $t=2$, $d_1=-2 d_2$ and
      $k_1=k_{\rm off}$. The values of $v_{\rm sat}$ and $\rm [ATP]_{1/2}$
      used in the normalization are those from the analytical model for all
      curves.}
    \label{fig:appendix}
  \end{figure}

\section{Conclusions}

With this simple model we could show that the collective motor velocity, for
example measured in a gliding assay, can show non-trivial dependence on the
ATP concentration. A careful examination of its limiting cases (low and high
[ATP]) can reveal details about the duty cycle that would otherwise require
single molecule measurements which are not only more demanding, but also more
affected by thermal noise. In particular, we show that a force-velocity
relationship that resembles the Michaelis-Menten shape (\ref{eq:michaelis})
indicates that the motor has at least 3 bound states and that ATP binding
occurs after the power stroke. We would expect this and largely find it for
muscle myosin
\cite{kron-spudich:86,Warshaw.Trybus1991,Debold.Warshaw2008,Debold.Walcott2011,Canepari.Bottinelli2012,Persson.Mansson2013}
and various dyneins \cite{Kagami.Kamiya1992,Lorch.Hunt2008}. A quadratic
dependence, on the other hand, is the signature of ATP binding before the
power stroke. This could apply to non-processive kinesin family motors, even
though some available diagrams \cite{deCastro.Stewart1999} do not yet show a
visible difference.

A particularly interesting aspect of the model is the theoretical possibility
that a motor could reverse its direction depending on the ATP
concentration. To our knowledge, no such behaviour has been reported in
natural motor proteins so far.  There are, however, kinesin-5 Cin8 that
switches direction depending on the ionic strength
\cite{Roostalu.Surrey2011,Gerson-Gurwitz.Gheber2011} and dynein that reverses
upon addition of phosphate \cite{Walter.Steffen2012}.  Notable achievements
involving artificial inversion and/or direction switching include an insertion
in the myosin lever arm that reversed its direction
\cite{Tsiavaliaris.Manstein2004} and a myosin construct that can switch
direction depending on the calcium concentration \cite{Chen.Bryant2012}.  The
possibilities to engineer motor molecules should eventually allow an
adjustment of lever displacements in individual states and creation of a motor
whose direction of motion would depend on the ATP concentration.

Finally, the same approach that we used here to describe the longitudinal
motion could also be used for rotational motion of filaments, driven by
lateral power-strokes in motor proteins. Filament rotation by non-processive
myosins can be caused by the fact that for steric reasons a myosin head can
only bind to certain ``target zones'' on an actin filament \cite{Vilfan2009b},
but in addition myosin can have an off-axis component of the power stroke.
The interplay between the two-contributions could lead to a cross-over from
left- to right handed rotation depending on the ATP concentration. In
addition, if a motor has several lateral power strokes, this alone could
already lead to complex dependencies of the helical pitch on the ATP
concentration. 

\acknowledgements 
This work was supported by Slovenian Research Agency (Grant
P1-0099 and J1-5437).
\vspace{1cm}

\appendix

\section{Simulation of finite ensembles}
\label{appendix}

In order to validate the steady state assumption on which the theory in this
paper is based, we show some simulation results on a finite ensemble of motors
in this Appendix. All results were obtained with a kinetic Monte Carlo
simulation (Gillespie algorithm) on a system of $n$ rigidly coupled motors
with the same properties as described in the main text. For the sake of
simplicity, all simulations were carried out with $\tau_{\rm det}=0$
(corresponding to $k_{\rm on}\to \infty$), which means that a group of $n=2$
motors is already processive (i.e., the two motors will not detach from the
track simultaneously). Figure \ref{fig:appendix} shows the velocities for the
same parameters as in Figs.\ \ref{figure4} and \ref{figure5}, but with finite
numbers ($n=2,5,10,20$) of motors. Although very small groups ($n=2$) show
some significant deviation, the analytical result becomes almost exact for
$n\gtrsim 10$ motors.

\end{document}